\documentclass[journal,onecolumn]{IEEEtran}

\usepackage{amsthm}
\usepackage{subfigure}
\usepackage{graphicx}
\usepackage{amssymb,amstext,amsmath}
\usepackage{float}
\usepackage{bbm}
\usepackage{bm}
\usepackage{multirow}
\usepackage{makecell}
\usepackage{color}
\usepackage{caption}
\usepackage{verbatim}
\usepackage{tikz}
\usepackage{algorithm}
\usepackage[noend]{algpseudocode}

\def\x{{\mathbf x}}

\newcommand{\normalized}{\widetilde}

\newcommand{\argmax}[1]{\underset{#1}{\arg\max}}

\newcommand\blfootnote[1]{%
  \begingroup
  \renewcommand\thefootnote{}\footnote{#1}%
  \addtocounter{footnote}{-1}%
  \endgroup
}

\pdfoutput=1
\begin{document}

\title{Heretical Multiple Importance Sampling}

\author{V\'ictor Elvira,~\IEEEmembership{Member,~IEEE,}
         Luca Martino,
        David Luengo,~\IEEEmembership{Member,~IEEE,}\\
        and~M\'onica F. Bugallo,~\IEEEmembership{Senior Member,~IEEE,}
}

%\markboth{IEEE SIGNAL PROCESSING LETTERS,~Vol.~X, No.~Y, DATE}%
%{Shell \MakeLowercase{\textit{et al.}}: Bare Demo of IEEEtran.cls for Journals}

\maketitle
\blfootnote{V. Elvira is with the {Department} of Signal Theory and Communications, Universidad Carlos III de Madrid,  Legan\'es (Spain).}
\blfootnote{ {L. Martino is with the Image Processing Laboratory, University of Valencia, Valencia (Spain).}}
\blfootnote{D. Luengo is with the {Department} of Signal Theory and Communications, Universidad Polit\'ecnica de Madrid, Madrid (Spain).}
\blfootnote{M. F. Bugallo is with the {Department} of Electrical and Computer Engineering, Stony Brook University, New York (USA).}
\blfootnote{This work has been supported by the Spanish government's projects TEC2013-41718-R and TEC2015-64835-C3-3-R; BBVA Foundation's project MG-FIAR; ERC grant 239784 and AoF grant 251170; NSF's Award CCF-0953316; and EU's Marie Curie ITN MLPM2012 (Ref. 316861).}

\begin{abstract}
Multiple Importance Sampling (MIS) methods approximate moments of complicated distributions by drawing samples from a set of proposal distributions. Several ways to compute the importance weights assigned to each sample have been recently proposed, with the so-called deterministic mixture (DM) weights providing the best performance in terms of variance, at the expense of an increase in the computational cost. A recent work has shown that it is possible to achieve a trade-off between variance reduction and computational effort by performing an \emph{a priori} random clustering of the proposals (partial DM algorithm). In this paper, we propose a novel ``heretical'' MIS framework, where the clustering is performed \emph{a posteriori} with the goal of reducing the variance of the importance sampling weights. This approach yields biased estimators with a potentially large reduction in variance. Numerical examples show that heretical MIS estimators can outperform, in terms of mean squared error (MSE), both the standard and the partial MIS estimators, achieving a performance close to that of DM with less computational cost.

\end{abstract}

%%%%%%%%%%%%%%%%%%%%%%%%%
%%%%%%%%%%%%%%%%%%%%%%%%%

%%%%%%%%%%%%%%%%%%%%%%%%%
%%%%%%%%%%%%%%%%%%%%%%%%%
\section{Introduction}
\label{sec_intro}
%%%%%%%%%%%%%%%%%%%%%%%%%
%%%%%%%%%%%%%%%%%%%%%%%%%

Importance sampling (IS) methods are widely used in signal processing to approximate statistical moments of {random variables ({r.v.}) by} a set of weighted samples. 
In particular, they are employed when the targeted probability density function (pdf) of the r.v. is complicated and its moments cannot be analytically computed, and/or drawing samples from {the targeted pdf} is impossible.
In {those cases}, samples are drawn from a simpler proposal pdf, and a set of weights associated to them are used to measure the discrepancy between the proposal and target densities \cite{Robert04,Liu04b}.
While the standard IS approach uses a single proposal density to draw all the samples, {multiple importance sampling} (MIS) schemes use a set of different proposal distributions to reduce the variance of the estimators \cite{Veach95,Owen00,elvira2015generalized}.
%
%The samples drawn from the different proposals in this also receive associated weights.
{Due} to the existence of multiple proposals, there are many different alternatives for the sampling and weighting procedures (see \cite{elvira2015generalized} for a thorough description of several valid schemes). 
In the {standard MIS} scheme, {the weight of a particular sample} is proportional to the ratio between the target pdf and the proposal that generated {that} sample, both evaluated at the sample value (see for instance \cite[Section 3.3.1.]{Robert04}).
The so-called {deterministic mixture} (DM) MIS scheme provides the best performance (in terms of variance of the associated estimators) of all the alternatives proposed so far (see \cite[Theorems 1 and 2]{elvira2015generalized}).
Nevertheless, this variance reduction is achieved at the expense of increasing the computational load w.r.t. the standard MIS scheme.

A {partial DM} MIS scheme, where proposals are {randomly clustered \emph{a priori}} into disjoint sets and the partial DM weight of each sample is computed {by} taking into account only the proposals of the specific subset, has recently been proposed \cite{elvira2015efficient}.
%
%\cblue{The} disjoint sets are constructed randomly and independently from the samples.
%
The variance and the computational complexity of this {partial DM} scheme falls in-between the standard and DM approaches, allowing the practitioner to select {the} desired computational {budget} in order to reduce the variance.
{Note that the} effect of using the DM or partial DM weights instead of the standard weights is linked to other approaches that attempt to reduce the {mean squared error (MSE)} of the IS estimators.
It is well known that the IS weights are usually (right) heavy-tailed distributed when the proposal and the target distribution have a mismatch \cite{ionides2008truncated, koblents2015population,vehtari2015pareto}, {thus resulting in unstable and high variance estimators.}
Several techniques address this issue by nonlinearly transforming the IS weights.
Illustrative {examples} are {the truncated IS} algorithm \cite{ionides2008truncated}, {clipping NPMC} \cite{koblents2015population}, or {Pareto-smoothed IS} \cite{vehtari2015pareto}.
All {of} these techniques attain a large reduction in the variance of the estimators in many useful scenarios at the expense of introducing bias.

In this paper, we present a novel {``heretical'' MIS} methodology that extends the {partial DM} framework of \cite{elvira2015efficient} with the goal of reducing the heavy tails of the distribution of weights.
The proposal densities are partitioned into disjoint subsets, as in {partial DM}.
However, unlike \cite{elvira2015efficient} (where the partition is fixed {\emph{a priori}}), here the partitioning is performed {\emph{a posteriori}} (i.e., once the samples have been drawn), focusing on finding configurations that reduce the largest weights. 
{This approach yields biased estimators with a potentially large reduction in variance, especially for small sample sizes \cite{eldar2008rethinking,kay2008rethinking}.
Numerical examples show that heretical MIS estimators can outperform, in terms of MSE, both the standard and {the partial} DM MIS estimators.}
%

%%%%%%%%%%%%%%%%%%%%%%%%%
%%%%%%%%%%%%%%%%%%%%%%%%%
\section{Importance Sampling}
\label{sec:is}
%%%%%%%%%%%%%%%%%%%%%%%%%
%%%%%%%%%%%%%%%%%%%%%%%%%

Let us denote the vector of unknown parameters to be inferred as $\x\in \mathcal{X}\subseteq \mathbb{R}^{d_x}$ with pdf
\begin{equation}
	\normalized{\pi}(\x)
		= \frac{\pi(\x)}{Z},
		\label{eq_posterior}
\end{equation}
where $\pi(\x)$ is the unnormalized target function, and $Z$ is the partition function.
The goal is computing some particular moment of $\x$, which can be defined as
\begin{equation}
	I =  \int_{\mathcal{X}} f(\x) \normalized{\pi}(\x) d\x = \frac{1}{Z} \int_{\mathcal{X}} f(\x) \pi(\x) d\x,
\label{eq_integral}
\end{equation}
where $f(\cdot)$ can be {any square-integrable function of $\x$ w.r.t. $\pi(\x)$}.
Unfortunately, in many cases an analytical solution of Eq. \eqref{eq_integral} cannot be computed, and Monte Carlo methods are used to approximate $I$.

%%%%%%%%%%%%%%%%%%%
\subsection{Importance sampling with a single proposal}
%%%%%%%%%%%%%%%%%%%

Let us consider $N$ samples, $\{\x_n\}_{n=1}^N$, drawn from a single proposal pdf, $q(\x)$, with heavier tails than the target, $\pi(\x)$.
Each sample has an associated importance weight given by
\begin{equation} 
	w_n= \frac{\pi(\x_n)}{{q(\x_n)}}, \quad n=1,\ldots,N.
\label{is_weights_static}
\end{equation} 
Using the samples and weights, the moment of interest can be approximated with a self-normalized estimator as
\begin{equation}
%	\tilde{I}_{\textrm{IS}} = \frac{1}{\sum_{j=1}^N w_j} \sum_{i=1}^N w_i  f(\x_i) = \frac{1}{N\hat{Z}}  \sum_{i=1}^N w_i f(\x_i),
	\tilde{I}_{\textrm{IS}} = \frac{1}{N\hat{Z}}  \sum_{n=1}^N w_n f(\x_n),
\label{eq_norm_est}
\end{equation}
where $\hat{Z}=\frac{1}{N}\sum_{n=1}^N w_n$ is an unbiased estimator of $Z=\int_{\mathcal{D}} \pi(\x) d\x$ \cite{Robert04}.
If the normalizing constant is known, then it is possible to use the unnormalized estimator, given by
\begin{equation}
	\hat{I}_{\textrm{IS}} = \frac{1}{NZ}  \sum_{n=1}^N w_n f(\x_n).
\label{eq_unnorm_est}
\end{equation}

Note that $\tilde{I}_{\textrm{IS}}$ is asymptotically unbiased, whereas $\hat{I}_{\textrm{IS}}$ is unbiased.
Both $\tilde{I}_{\textrm{IS}}$ and $\hat{I}_{\textrm{IS}}$ are consistent estimators of $I$ and their variance is directly related to the discrepancy between $\normalized\pi(\x)|f(\x)|$ and $q(\x)$  \cite{Robert04, kahn1953methods}.
{However, when} several different moments $f$ must {be} estimated or the function $f$ is unknown \emph{a priori}, a common strategy is decreasing the mismatch between the proposal $q(\x)$ and the target $\normalized \pi(\x)$ \cite[Section 3.2]{doucet2009tutorial}, which is equivalent to minimizing the variance of the weights (and also the variance of the estimator $\hat{Z}$). %The usual approach for increasing the performance of IS methods is selecting de proposal $q(\x)$ adequately in order to reduce that discrepancy. 

%%%%%%%%%%%%%%%%%%%
%\subsection{Performance: the curse of heavy tailed distributed weights}
%%%%%%%%%%%%%%%%%%%
 
%%%%
\subsection{Multiple importance sampling (MIS)}
\label{sec:mis}
%%%%
%\subsection{Estimators and importance sampling weights}
%%%%%%%%%%%%

%It is well known that the variance of IS estimators is directly related to the discrepancy between $\pi(\x)|f(\x)|$ and $q(\x)$ \cite{Robert04, kahn1953methods}.
%
Since the target density can only be evaluated point-wise, but cannot be easily characterized in many cases, finding a single good proposal pdf, $q(\x)$, is not always possible. A robust alternative is then using a set of proposal pdfs, $\{q_n(\x)\}_{n=1}^N$ \cite{Veach95,Owen00}.
This scheme is usually known in the literature as {multiple importance sampling (MIS)}, and it is an important feature of many state-of-the-art adaptive {IS} algorithms (e.g., \cite{Cappe04, CORNUET12, APIS15}).

%\subsection{Common Multiple Importance Sampling approaches}

A general MIS framework has recently been proposed in \cite{elvira2015generalized}, where several sampling and weighting schemes can be used. Here, we briefly review the most common sampling and two common weighting schemes. Let us draw one sample from each proposal pdf, i.e., 
\begin{equation} 
\x_n \sim q_n(\x), \qquad n=1,...,N. \label{eq_sampling_mis}
\end{equation}
The most common weighting strategies in the literature are:
\begin{enumerate}
\item \mbox{{\it Standard MIS} (s-MIS) \cite{Cappe04}:}
\begin{equation} 
w_n= \frac{\pi(\x_n)}{{q_n(\x_n)}}, \quad n=1,\ldots, N. \label{eq_w_smis}
\end{equation}
\item {\it Deterministic mixture MIS} (DM) \cite{Owen00}:
\begin{equation} 
	w_n=\frac{\pi(\x_n)}{\psi(\x_n)}=\frac{\pi(\x_n)}{\frac{1}{N}\sum_{j=1}^{N}q_j(\x_n)}, \quad n=1,\ldots, N,
	\label{eq_w_dmmis}
\end{equation}
where $\psi(\x)=\frac{1}{N}\sum_{j=1}^{N}q_j(\x)$ is the mixture pdf, composed of all the proposal pdfs, evaluated at $\x$.
\end{enumerate}

%Note that, while s-MIS interprets that the $\x_n$ are realizations of the proposal $q_n$, DM-MIS uses the whole mixture $\psi(\x)$ as the weight in the denominator for all the samples.
%
Once more, the estimator of Eq. \eqref{eq_norm_est} using any of both sets of weights {(i.e., \eqref{eq_w_smis} or \eqref{eq_w_dmmis})} is consistent and asymptotically unbiased, whereas the estimator of Eq. \eqref{eq_unnorm_est} is both consistent and unbiased.
%\footnote{There are several interpretations to explain why DM-MIS yields a valid estimator. For instance, the $N$ samples can be considered samples drawn from $\psi(\x)$ with a variance reduction technique such an extended stratified sampling procedure \cite{OwenBook}.}
%
The superiority of the DM approach w.r.t. the s-MIS,  in terms of variance of the estimator $\hat{I}$, has been proved in \cite{elvira2015generalized}.
However, although both alternatives perform the same number of target evaluations, the DM estimator is computationally more expensive w.r.t. the number of proposal evaluations. In particular, s-MIS and DM require $N$ and $N^2$ evaluations, respectively.
In some scenarios (e.g., with a large number of proposals $N$) the extra number of proposal evaluations can be a major issue, and alternative efficient solutions have to be {devised}.

%%%%
\subsection{Partial Multiple Importance Sampling (p-DM)}
%%%%
A new general framework for {building estimators}, called {partial} DM {and {characterized by} a performance and computational complexity in between s-MIS and DM}, was introduced in \cite{elvira2015efficient}.
Let us assume that we draw $N$ samples using Eq. \eqref{eq_sampling_mis}, and a partition of all {the} $N$ proposals into $P$ disjoints subsets of $M$ proposals ($N=PM$) is defined \emph{a priori}. More precisely, the set of indices $\{1,\ldots,N\}$ is partitioned into $P$ disjoint subsets of $M$ indices, $\mathcal{S}_p$ with $p=1,\ldots,P$, s.t.
\begin{equation}
	\mathcal{S}_1 \cup \mathcal{S}_2\cup \ldots \cup \mathcal{S}_P= \{1,\ldots,N\},
\end{equation}
where $\mathcal{S}_k \cap \mathcal{S}_\ell = \emptyset$ for all $k,\ell=1,\ldots,P$ and $k\neq \ell$.
Each subset, $\mathcal{S}_p = \{j_{p,1}, j_{p,2},\ldots, j_{p,M}\}$, contains $M$ indices, $j_{p,m}\in \{1,\ldots,N\}$ for $m=1,\ldots,M$ and $p=1,\ldots, P$,\footnote{For the sake of simplicity, and without loss of generality, it is assumed that all {the} subsets have the same number of proposals.} and then each sample $\x_n$ receives a ``partial'' DM weight considering at its denominator the mixture of all the proposals of the subset.
Following this strategy, the weights of the samples of the $p$-th mixture are computed as
\begin{equation}
	w_n = \frac{\pi(\x_{n})}{\psi_p(\x_{n})} 
		= \frac{\pi(\x_{n})}{\frac{1}{M}\sum_{j\in\mathcal{S}_p}q_{{j}}(\x_{n})}, \quad n \in\mathcal{S}_p.
\label{p_dm_weights_static}
\end{equation}

It can be proved that the p-DM approach yields consistent estimators, with a performance and computational complexity in between s-MIS and DM (see \cite[Appendix B]{elvira2015efficient}), i.e.,
\begin{equation}
	\textrm{Var}(\hat{I}_{\textrm{DM}}) \le \textrm{Var}(\hat{I}_{\textrm{p-DM}}) \le \textrm{Var}(\hat{I}_{\textrm{s-MIS}}).
\label{eq_three_variances_comparison}
\end{equation}
Moreover, in p-DM the number of evaluations of the proposal pdfs is $PM^2$.
Since $N \leq PM^2 = NM \leq N^2$, the computational cost is larger than that of the s-MIS approach ($M$ times larger), but lower than that of the f-DM approach (since $M\leq N$). 
{The} particular cases $P=1$ and $P=N$ correspond to the DM and the s-MIS approaches, respectively.

%%%%%%%%%%%%%%%%%%%%%%%%%
%%%%%%%%%%%%%%%%%%%%%%%%%
\section{Heretical MIS}
\label{sec:heretic}
%%%%%%%%%%%%%%%%%%%%%%%%%
%%%%%%%%%%%%%%%%%%%%%%%%%
In adaptive {IS} algorithms, the proposals are iteratively adapted in order to {reduce} {the mismatch w.r.t. the target} (see for instance PMC \cite{Cappe04}, AMIS \cite{CORNUET12}, or  APIS  \cite{APIS15}).
However, {this} does not always occur, and {a substantial mismatch can still remain in the first iterations} even when the adaptation is successful (see for instance the discussion in \cite[Section 4.2.]{koblents2015population}). 
This mismatch yields right heavy {tailed} distributed weights; extremely large weights can be associated to samples drawn in the tails of the proposals {for which} the target evaluation is high.
In this situation, the estimator {can be dominated in practice by a single} sample.
This is a well known fact that can result in a large increase in the variance of the estimators, even yielding estimators with infinite variance (see for instance \cite{Owen00}).
This example intuitively explains how DM reduces the variance of the weights: in the aforementioned situation, if there is another proposal in the neighborhood of $\x_n$, the evaluation of the mixture proposal is no longer too small, and the weight does not take an extremely large value.
This will be exploited by the proposed algorithm in order to reduce the variance of the {MIS} {weights}.

\subsection{Proposed algorithm}
\label{sec:algorithm}

In the {novel heretical DM (h-DM) MIS} algorithm, the proposals are clustered once the samples $\{\x_n\}_{n=1}^N$ have been drawn with the goal of reducing the largest weights, {unlike in the p-DM algorithm where the clustering is performed \emph{a priori}}.
The {proposed approach} is summarized in Algorithm \ref{alg_hMIS}.
The sampling (line 1) and the {initial} weighting (line 2) are {performed} exactly as in s-MIS.
Afterwards, the algorithm selects the largest weight, $w_{n^*}$ (line 5), and the proposal $j^*$ that maximizes $q_{j^*}(\x_{n^*})$.
Then, $q_{n^*}$ and $q_{j^*}$ are clustered into the same subset (lines 7--13).
This process is repeated until all {the} proposals have been allocated to one of all the $P$ subsets (or until some other rule is fulfilled, as discussed in {Section \ref{sec_disc_extension}}).
Finally, the weights are recomputed according to the p-DM weights of Eq. \eqref{p_dm_weights_static} (lines 17--18).

\subsection{{Computational complexity and performance}}
\label{sec:comp}

{The h-DM algorithm performs $N^2/P \leq N^2$ proposal evaluations, thus reducing the computational complexity of the full DM approach, which is now comparable to that of the p-DM algorithm.
%
%Therefore, the h-DM is light in terms of calculations, {with} 
%
The most costly step (w.r.t. the p-DM) is the search for the proposal $j^*$ that maximizes the evaluation $q_{j^*}(\x_{n^*})$ (line 6).}
However, let us remark that the computational cost of this stage can be substantially alleviated.
Firstly, a smart search  can be performed in order to avoid {checking} all the available proposals.
Secondly, suboptimal search strategies are likely to work similarly well; the variance of the weights {can still be} notably reduced if close enough proposals (not necessarily the closest {ones}) to the samples with large weights are found.
Thirdly, in order to find the closest proposal, $q_{j^*}$, it is not always necessary to compute the whole evaluation.
Finally, the (partial) evaluation of some of the proposals can be reused at the reweighting stage (line 17).
{Similarly to the p-DM, the performance of h-DM improves when increasing $P$. In addition, for a specific choice of $P$, h-DM reduces the variance of the weights at the expense of introducing bias. While a theoretical analysis is still required, the variance reduction can be large, especially for low values of $P$. This explains why h-DM clearly outperforms p-DM in the experiments of Section \ref{sec:results}.}

%% Algorithm
\begin{algorithm}
\caption{Heretical {MIS}}\label{alg_hMIS}
\begin{algorithmic}[1]
%\Procedure{MyProcedure}{}
\State Draw $N$ samples, one from each proposal pdf,
$$\x_n \sim q_n(\x), \qquad n=1,...,N.$$
\State Compute the standard IS weights 
$$w_n= \frac{\pi(\x_n)}{q_n(\x_n)}, \qquad n=1,...,N.$$
\State Initialize the subsets $\mathcal{R} = \{1,...,N \}$, $\mathcal{A} = \{1,...,N \},$
and $\mathcal{S}_p = \{\emptyset\}$ with $p=1,...,P.$
\While {$\mathcal{R} \neq \emptyset$}
\State Find the index of the maximum weight:
	\begin{equation*}
		n^* = \argmax{n \in \mathcal{R}} \; w_n
	\end{equation*}
\State Find the index of the {closest} proposal to $\x_{n^*}$: %in terms of Mahalanobis distance.
%$$j^* ={ \arg\min \limits}_{j \in \{1,...,N\} \backslash n^*} \; (\x_{n^*} - \bmu_j)^{\top} \bSigma _j^{-1}(\x_{n^*} - \bmu_j)$$
%$$j^* ={ \arg\min \limits}_{j \in \mathcal{A}} \; (\x_{n^*} - \bmu_j)^{\top} \bSigma _j^{-1}(\x_{n^*} - \bmu_j)$$
	\begin{equation*}
		j^* = \argmax{j \in \mathcal{A}} \; q_j(\x_{n^*})
	\end{equation*}
%\State $i \gets \textit{patlen}$
%\BState \emph{top}:
%\BState \emph{loop}:
\If {$j^*$ {belongs} already {to} a subset $\mathcal{S}_{m_{j^*}}$}
\State cluster $n^*$ {into} that subset: $\mathcal{S}_{m_{j^*}} = \mathcal{S}_{m_{j^*}} \cup n^*$
\State save the subset where $n^*$ is stored: $m_{n^*} = m_{j^*}$
\Else
\State cluster $n^*$ and $j^*$ together:\footnotemark $\mathcal{S}_{m} = \mathcal{S}_{m}  \cup n^*  \cup j^* $
\State save the subset where $n^*$ is stored: $m_{n^*} = m$
\State save the subset where $j^*$ is stored: $m_{j^*} = m$
	
%i.e. samples $\x_{n^*}$ and $\x_{j^*}$ will have the same denominator function.
\EndIf 
\State $\mathcal{R} =  \mathcal{R}  \backslash n^*$
\If{$|\mathcal{S}_{n^*}|=M$}
\State $\mathcal{A}$ = $\mathcal{A} \backslash \mathcal{S}_{n^*}$. 
\EndIf 
%\If{$\mathcal{R} \neq \emptyset$}
%\EndIf 
\EndWhile
\State Recompute the weights $w_n$ using Eq. \eqref{p_dm_weights_static}
\State Compute the normalized weights as $\bar w_n = \frac{w_n}{\sum_{k=1}^N w_n}$
%\EndProcedure
\end{algorithmic}
%\label{alg_hmis}
%\caption{Heretical Multiple Importance Sampling}
\end{algorithm}
\footnotetext{In this case, $n^*$ and $j^*$ are stored in any $\mathcal{S}_m$, with the condition that this subset has two available slots, i.e., $|\mathcal{S}_m|\leq M - 2$. If none of the subsets has two available slots, $n^*$ and $j^*$ are randomly allocated. Note that more advanced strategies can be developed in this situation, but we keep the algorithm simple for {the} sake of clarity in the explanation.}

\vspace{-5mm}

\subsection{Extensions}
\label{sec_disc_extension}

The proposed algorithm opens the door to a novel class of MIS algorithms with one feature in common: constructing the weighting functions after drawing the samples (i.e., the weighting function itself is a r.v.). 
%
%We recall that the proposed algorithm is a proof of concept, showing the performance improvement and it is justified with the goal of reducing the variance of the weights. 
%
Several aspects of the {core} algorithm {proposed here} can be improved: the computational complexity (related to the number of subsets $P$) might be {automatically determined} by the algorithm, the subsets might have different number of elements, or the partitioning algorithm could be stopped once the largest weights have been reduced (performing the rest of the allocation randomly).
{For instance, we propose a simple extension of the clustering method by introducing a parameter ($0 \leq\alpha \leq 1$) that represents the fraction of proposals that are clustered following Algorithm \ref{alg_hMIS}.
Whenever more than $\alpha N$ proposals have been allocated, the remaining ones are randomly clustered as in p-DM.
Hence, $\alpha=0$ corresponds to the p-DM in \cite{elvira2015efficient}, while $\alpha=1$ corresponds to Algorithm \ref{alg_hMIS}.} 
{The extension of the algorithm proposed in this section largely reduces the complexity of the clustering algorithm (roughly by a fraction of $\alpha$).}

%%%%
\subsection{Connection with other techniques}
%%%%

Several algorithms in the literature propose alternative ways to reduce the variance of the IS weights.
The common approach consists {of} {clipping} the largest unnormalized weights to a certain value (typically, the minimum of the set of transformed/clipped weights), and then normalizing the weights. 
%Notable examples of this approach can be found in \cite{ionides2008truncated, koblents2015population,vehtari2015pareto}. 
In \cite{ionides2008truncated} and \cite{koblents2015population}, a fraction of the largest unnormalized weights are clipped to the minimum of those weights.
%
%Solid proofs of asymptotical consistency of the estimators are provided in \cite{koblents2015population}.
%
The method proposed in \cite{vehtari2015pareto} is an advanced implementation of \cite{ionides2008truncated}, where the distribution of the weights is fitted by a Pareto distribution.

{In h-DM, the largest unnormalized weights are also attenuated, but the approach followed is completely different, since the procedure is based on the construction of partial MIS estimators.
The statistical meaning of the h-DM weights is then supported by the {DM} interpretation, present in DM or p-DM.
While \cite{ionides2008truncated}, \cite{koblents2015population} and \cite{vehtari2015pareto} introduce bias due to the {nonlinear} transformation of the weights, in h-DM the bias appears because the values of the samples are used to perform the clustering.
A detailed theoretical study of the bias and consistency of h-DM is thus required.
In future works we plan to address this issue by adapting and extending the analysis techniques already developed in \cite{ionides2008truncated, koblents2015population,vehtari2015pareto}.}

%%%%%%%%%%%%%%%%%%%%%%%%%
%%%%%%%%%%%%%%%%%%%%%%%%%
\section{Numerical results}
\label{sec:results}
%%%%%%%%%%%%%%%%%%%%%%%%%
%%%%%%%%%%%%%%%%%%%%%%%%%

%\subsection{Example 1}
\textbf{Example 1}: Let us consider a unidimensional target which is itself a mixture of two Gaussian pdfs:
\begin{equation}
\label{Target2}
\pi(x)= \frac{1}{2}\mathcal{N}(x;a_1,c^2) + \frac{1}{2}\mathcal{N}(x;a_2,c^2),
\end{equation}
with $a_1=-3$, $a_2=5$, and $c^2=1$.

The goal is to approximate, via Monte Carlo, the expected value of $x\sim \pi(x)$, i.e., $E[x]=\int_{\mathbb{R}} x \pi(x) dx$ and $Z=\int_{\mathbb{R}} \pi(x) dx$.
We apply the MIS algorithms in a setup with $N=32$ Gaussian proposal pdfs, $\{q_{n}(x)=\mathcal{N}(x;{\mu}_{n},\sigma^2)\}_{n=1}^N$, where the means are equidistant in the interval $[-8,8]$, while their variances are all equal to $\sigma_n^2=3$ for $n=1,...,N$, {and} $L=32\cdot k$ samples with $k\in \{1,..,5�\}$, i.e., exactly $k$ samples per proposal {are drawn}.
We compare the performance of s-MIS, DM, p-DM, and the proposed h-DM.
{In} the last two algorithms {(p-DM and h-DM)} we set $P=16$, i.e., in p-DM the proposals are {randomly} clustered into $16$ subsets, {whereas} the allocation in h-DM is performed using Algorithm \ref{alg_hMIS}.
{Note that, for one-dimensional Gaussian proposals with equal variance, the maximization in line 6 of Algorithm \ref{alg_hMIS} is equivalent to
\begin{equation*}
	j^* = \argmax{j \in \mathcal{A}} \ (x_{n^*}-\mu_j)^2 = \argmax{j \in \mathcal{A}} \ |x_{n^*}-\mu_j|,
\end{equation*}
which requires only a partial evaluation of the proposals.%, as discussed in Section \ref{sec:algorithm}.}

Figure \ref{fig_temp} shows the performance of the different algorithms in terms of the MSE of the self-normalized estimator $\tilde I$ of the mean of the target distribution. It can be seen that h-DM outperforms p-DM in this scenario, obtaining a performance close to that of DM with a reduced computational cost.

\begin{figure}[htb]
  \centering
  \centerline{\includegraphics[width=10.5cm]{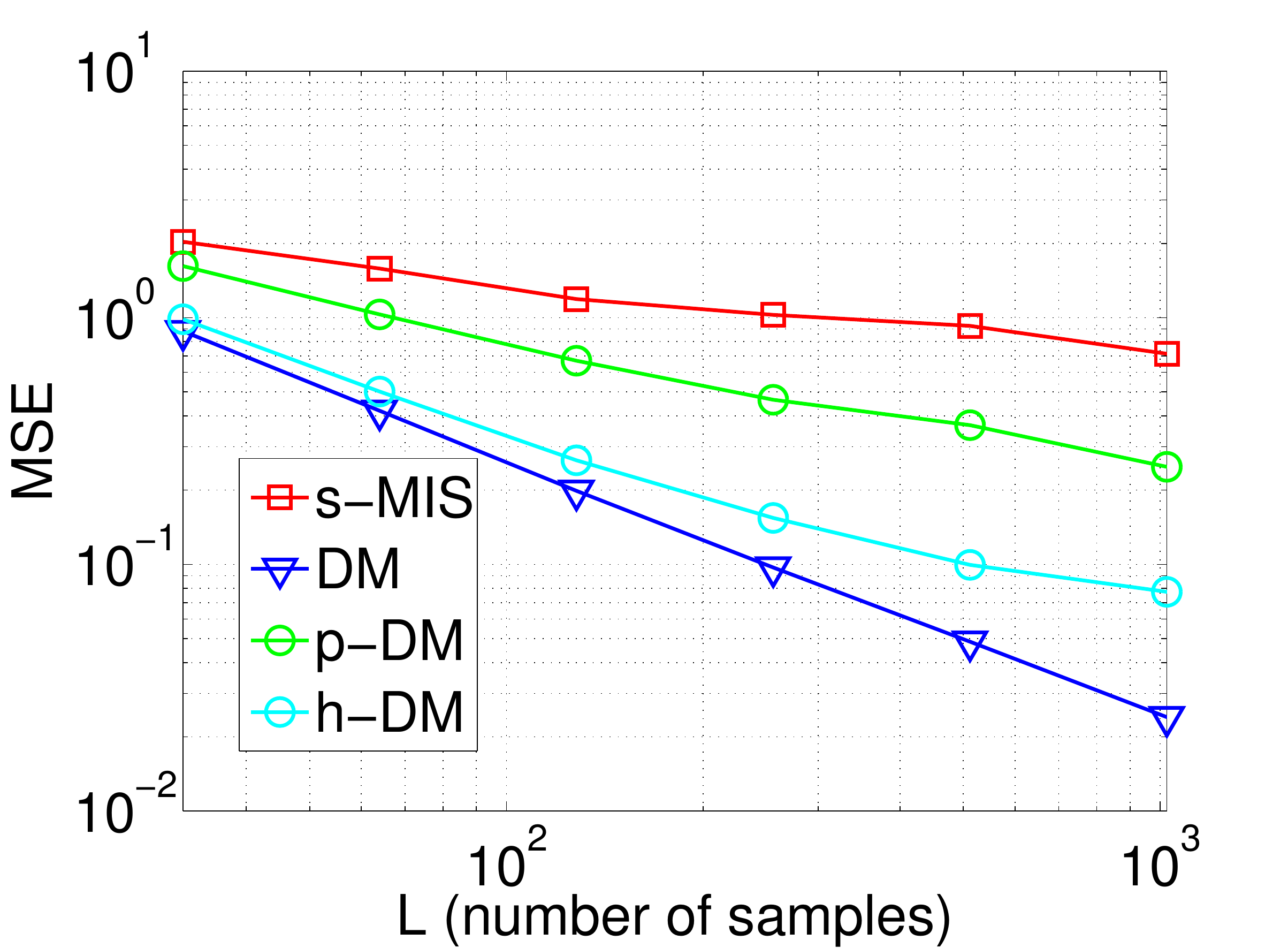}}
\caption{[{Ex.} 1] MSE of the self-normalized estimator $\tilde I$ of the mean of the target {using} $L=32\cdot k$ samples {for} $k\in \{1,..,5�\}$.}
\label{fig_temp}
\end{figure}

%%%%
%\subsection{Example 2}
%%%%

\textbf{Example 2}: Let us consider a unidimensional target which is a (normalized) mixture of five non-standardized {Student's t pdfs} with location parameters $a_1 =-3$,  $a_2=-1$, $a_3= 0$, $a_4= 3$ and $a_5 = 4$, scale parameters $c_i^2 = 1$ {and} degrees of freedom $\xi_i=5$, {both} for $i=1,...,5$.
The goal is to approximate the mean of the target.
We apply the MIS algorithm in a setup with $N=32$ non-standardized {Student's t pdfs}, where the location parameters $\mu_n$ are equidistant in the interval $[-8,8]$, while their scale parameters and degrees of freedom are equal to $\sigma_n^2=3$ and $\nu_n=4$ for $n=1,...,N$, respectively.
We compare the {p-DM} and the proposed h-DM, testing $P\in\{1,2,4,8,16,32�\}$, and thus $M\in\{32,16,8,4,2,1\}$.
Note that s-MIS corresponds to $P=32$ and $M=1$, whereas DM is obtained with $P=1$ and $M=32$.
{We use the modified algorithm suggested in Section \ref{sec_disc_extension} with $\alpha = 0.1$.}
Figure \ref{fig_ex2} shows the MSE of the unnormalized estimator $\tilde I$ of the mean of the target distribution.
Once more, h-DM outperforms p-DM {substantially (especially for $M<8$)}.

\begin{figure}[htp]
\centering
\subfigure[]{
\includegraphics[width=0.46\columnwidth]{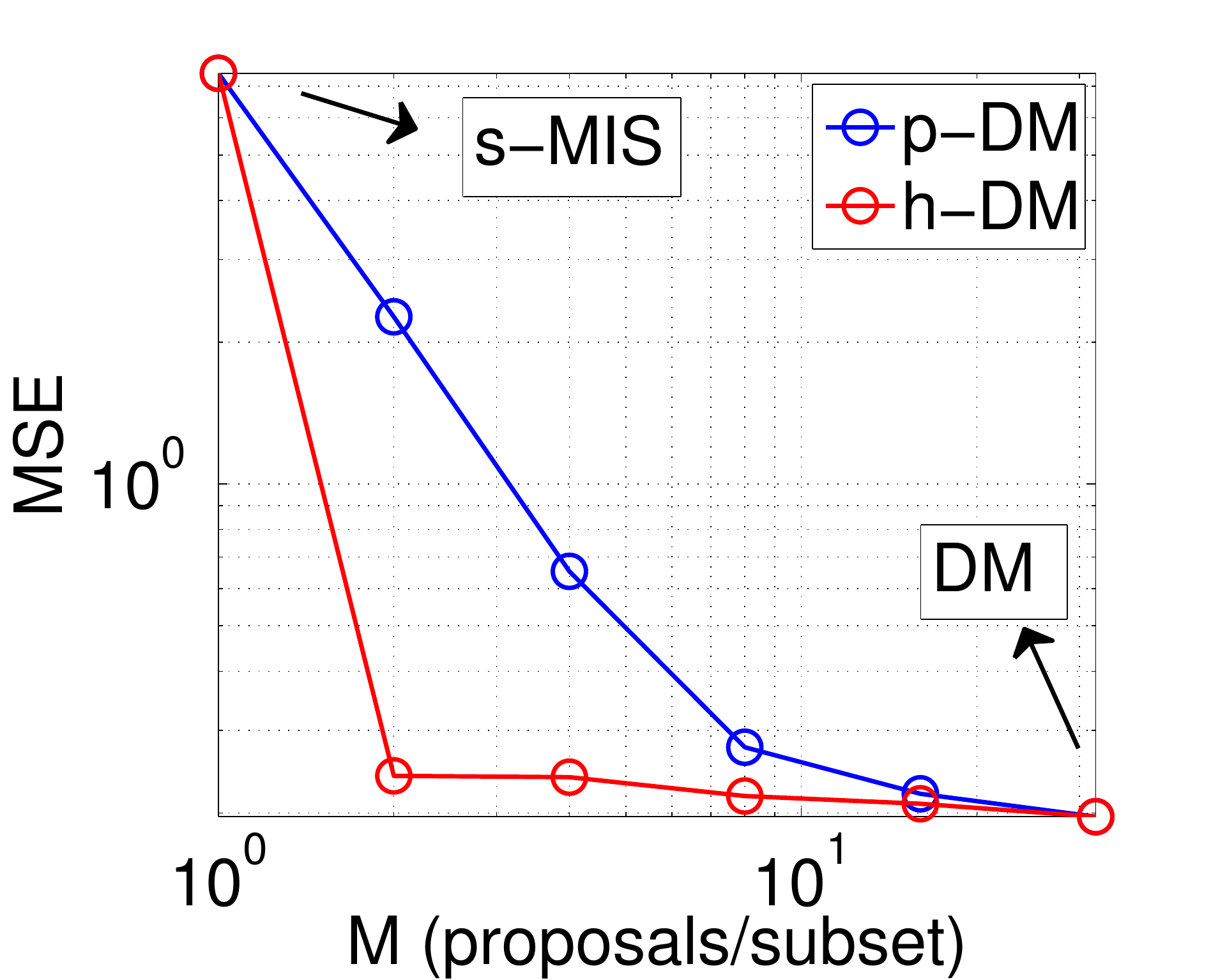}
}
\subfigure[]{
\includegraphics[width=0.46\columnwidth]{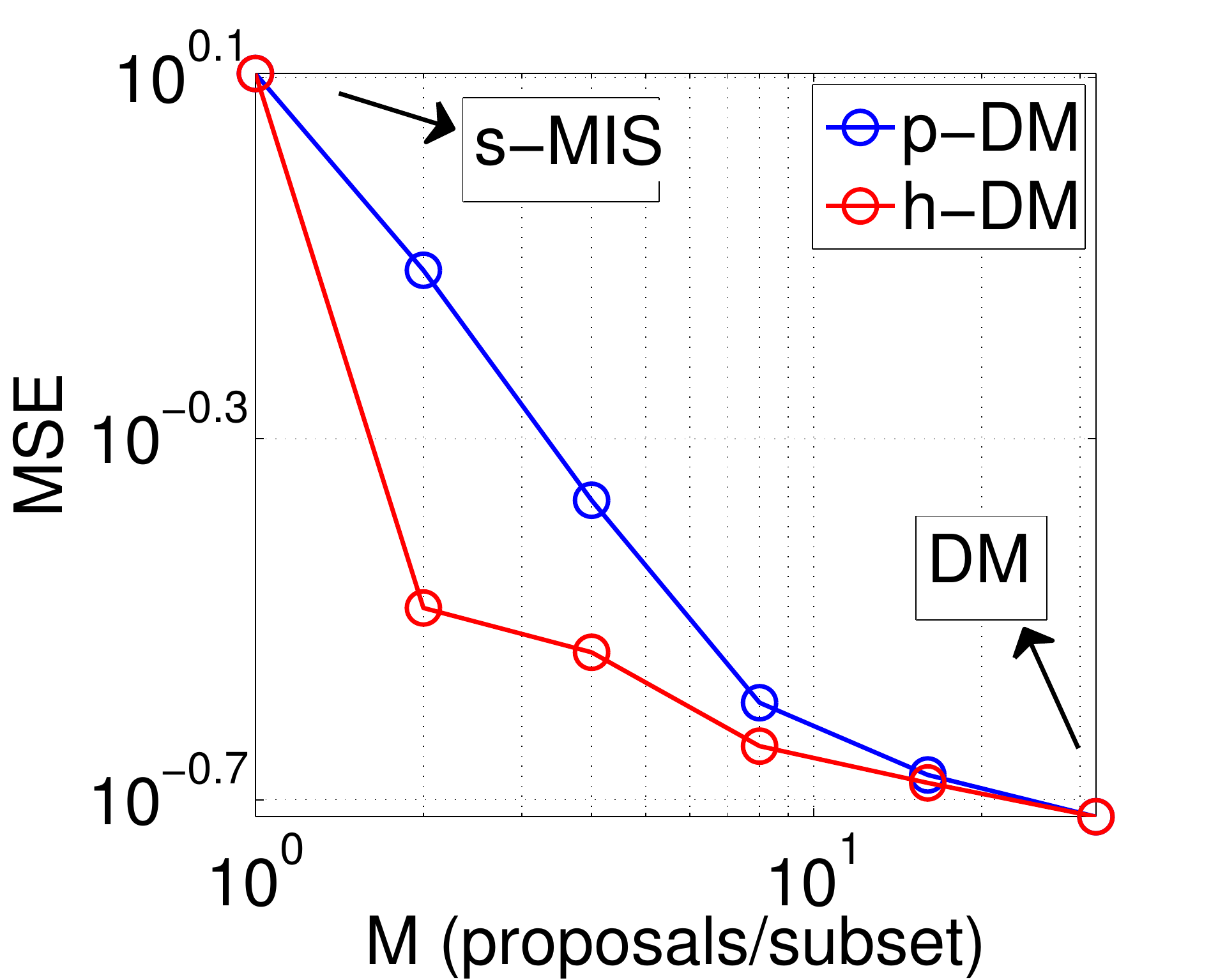}
} 
\caption{{[{Ex.} 2] } MSE of (a) $\hat I$ and (b) $\tilde I$.} %in the estimation of the mean of the target {using $M\in\{1,2,4,8,16,32\}$.}}
\label{fig_ex2}
\end{figure}

\vspace{-2mm}

%%%%%%%%%%%%%%%%%%%%%%%%%
%%%%%%%%%%%%%%%%%%%%%%%%%
\section{Conclusions}
\label{sec:conclusions}
%%%%%%%%%%%%%%%%%%%%%%%%%
%%%%%%%%%%%%%%%%%%%%%%%%%

{In this paper, we have proposed a heretical MIS framework that allows us to achieve a tradeoff between performance and computational cost by clustering the proposals \emph{a posteriori}.
The resulting estimators are biased, but their reduced variance can lead to a substantial decrease in MSE, especially for small sample sizes.
Future lines include developing more efficient clustering methods and performing a rigorous theoretical study of the bias and consistency of the estimators.}

\bibliographystyle{IEEEbib}

%\bibliography{bibliografia}

\end{document}